\documentstyle[times,pramana,epsf,epsfig,floats,amsmath,axodraw]{ias}

\input paperdef

\begin{document}



\thispagestyle{empty}
\setcounter{page}{0}
\def\thefootnote{\fnsymbol{footnote}}

\begin{flushright}
DCPT/06/108\\
IPPP/06/54\\
MPP--2006--94\\
hep-ph/0611371\\
\end{flushright}

\vspace{1cm}

\begin{center}

{\large\sc {\bf Testing the MSSM with the Mass of the \boldmath{$W$} Boson}}
\footnote{talk given by S.~Heinemeyer at the {\em LCWS06}, 9-13 March
  2006, Bangalore, India}

\vspace{1cm}

{\sc S.~Heinemeyer$^{1\,}$%
\footnote{
email: Sven.Heinemeyer@cern.ch
}%
, W.~Hollik$^{\,2}$%
\footnote{
email: hollik@mppmu.mpg.de
}%
, D.~St\"ockinger$^{\,3}$%
\footnote{
email: Dominik.Stockinger@durham.ac.uk
}%
,\\[.5em] A.M.~Weber$^{\,2}$%
\footnote{
email: Arne.Weber@mppmu.mpg.de
}%
 and G.~Weiglein$^{\,3}$%
\footnote{
email: Georg.Weiglein@durham.ac.uk
}%
}

\vspace*{1cm}

$^1$ Instituto de Fisica de Cantabria (CSIC--UC), Santander,  Spain 

\vspace*{0.4cm}

$^2$Max-Planck-Institut f\"ur Physik (Werner-Heisenberg-Institut),\\ 
F\"ohringer Ring 6, D--80805 Munich, Germany 

\vspace*{0.4cm}

$^3$IPPP, University of Durham, Durham DH1 3LE, U.K.

\end{center}

\vspace*{0.2cm}

\BC {\bf Abstract} \EC
We review the currently most accurate evaluation of the $W$~boson
mass, $\MW$, in the Minimal Supersymmetric Standard Model (MSSM). 
It consists of a full one-loop calculation, including the complex
phase dependence, all available
MSSM two-loop corrections as well as the full Standard Model result.
We analyse the impact of the phases in the scalar quark sector
on $\MW$ and compare the prediction for $\MW$ based on all known
higher-order contributions with the experimental results.

\def\thefootnote{\arabic{footnote}}
\setcounter{footnote}{0}

\newpage


\title{Testing the MSSM with the Mass of the \boldmath{$W$} Boson}

\author{S. Heinemeyer$^1$, 
        W.~Hollik$^2$, 
        D.~St\"ockinger$^3$, 
        A.M.~Weber$^2$, 
        G.~Weiglein$^3$}
\address{$^1$Instituto de Fisica de Cantabria (CSIC-UC), Santander, Spain\\
         $^2$ Max-Planck-Institut f\"ur Physik, 
              F\"ohringer Ring 6, D--80805 Munich, Germany\\
         $^3$ IPPP, University of Durham, Durham DH1 3LE, U.K.}

\keywords{MSSM, $W$ boson, precision observables}

\pacs{2.0}

\abstract{
We review the currently most accurate evaluation of the $W$~boson
mass, $\MW$, in the Minimal Supersymmetric Standard Model (MSSM). 
It consists of a full one-loop calculation, including the complex
phase dependence, all available
MSSM two-loop corrections as well as the full Standard Model result.
We analyse the impact of the phases in the scalar quark sector
on $\MW$ and compare the prediction for $\MW$ based on all known
higher-order contributions with the experimental results.

}

\maketitle


\section{Introduction}

The relation between the massive gauge-boson masses, $\MW$ and $\MZ$, in
terms of the Fermi constant, $\gf$, and the fine structure constant
$\al$, is of central importance for testing the electroweak theory:
\begin{equation}
\frac{G_{\mu}}{\sqrt{2}}=\frac{e^2}
      {8 \left(1-\frac{\MW^2}{\MZ^2}\right) \MW^2}\, (1+\De r)~.
\label{GFdeltarrelation}
\end{equation}
It is usually employed for predicting $\MW$ in the model under
consideration, where the loop corrections entery via $\De r$. 
This prediction can then be compared with the
corresponding experimental value. The current experimental accuracy for
$\MW$, obtained at LEP and the Tevatron, is $\de \MW = 29 \mev$
(0.04\%)~\cite{LEPEWWG,LEPEWWG2}.  
This experimental resolution provides a high sensitivity to quantum
effects involving the whole structure of a given model. The $\MW$--$\MZ$
interdependence is therefore an important tool for discriminating
between the Standard Model (SM) and its minimal supersymmetric
extension (MSSM)~\cite{susy}, see \citere{PomssmRep} for a recent review. 
Within the MSSM the $W$~boson mass, supplemented with other
electroweak precision observables, exhibits a
certain preference for a relatively low scale of supersymmetric
particles, see e.g.\ \citeres{ehow3,ehow4}. 
Consequently, a precise theoretical prediction for $\MW$ in terms of
the model parameters is of utmost importance for present and future
electroweak precision tests. 
A precise prediction for $\MW$ in the MSSM is also needed as 
a part of the ``SPA Convention and Project'', see \citere{spa}.

In \citere{MWweber} the currently most up-to-date evaluation of $\MW$
(i.e.\ $\De r$) in the MSSM has been presented. It consists of the
full one-loop 
calculation, taking into account the complex phase dependence (the
phases had been neglected so far in all previous calculations), the
full SM result~\cite{MWSM} and all available MSSM two-loop
contributions~\cite{dr2lA,dr2lB,drMSSMal2B,drMSSMal2A}. 
The corresponding Fortran program for the calculation of precision
observables within the MSSM will be made publicly avaliable \cite{WaZOb}.

In the numerical analysis below, exept for the parameter scan in
\refse{sec:scan},  for
simplicity we choose all soft 
SUSY-breaking parameters in the diagonal entries of the sfermion mass
matrices to be equal ($\equiv \msusy$).
In the neutralino sector the GUT relation 
$M_1 = 5/3 \, s_{\mathrm w}^2/c_{\mathrm w}^2 \, M_2$ has been used
(for real values). 
We have fixed the SM input parameters as%
\footnote{
Using the most up-to-date value for the top quark mass, 
$\mt = 171.4 \gev$~\cite{newestmt}, would lead to slightly lower
absolute $\MW$ values, while the impact on $\de\MW$ (see below) is
negligible. 
}%
\BE 
\begin{aligned}
G_{\mu} &= 1.16637\times 10^{-5}, &
\MZ &= 91.1875 \gev, & 
\als(\MZ) &= 0.117 , \\
\alpha &= 1/137.03599911, &
\De \al^{(5)}_{\textup{had}} &= 0.02761, & 
\De \al_{\textup{lep}} &= 0.031498, \\
\mt &= 172.5 \gev \textup{\cite{newmt}} \, &
\mb &= 4.7 \gev, &
m_\tau &= m_c = \ldots = 0 
\end{aligned} 
\label{eq:inputpars}
\EE
The complex phases appearing in the MSSM are experimentally 
constrained by their contribution to low energy observables such as
electric dipole moments (see \citere{PomssmRep} and references therin).
Accordingly (using the convention that $\phi_{M_2} =0$), in particular
the phase $\phi_\mu$ is tightly constrained~\cite{plehnix}, 
while the bounds on the phases of the third generation
trilinear couplings are much weaker.
The Higgs sector parameters are obtained from
the program \fhtt~\cite{feynhiggs}.


\section{Dependence on the complex phases in the squark sector}

Here we show the dependence of $\MW$ on the phases of the scalar quark
sector. The physical phases are $\phi_{\At} + \phi_\mu$ and
$\phi_{\Ab} + \phi_\mu$, where $A_{t,b}$ are the trilinear
Higgs-$\Stop,\Sbot$ coupling and $\mu$ is the Higgs mixing parameter.
We focus here on the mass shift $\de\MW$ arising from changing $\De r$
by the amount  
$\De r^{\rm SUSY}$, 
\begin{equation}
\de \MW = -\frac{\MW^{\rm ref}}{2}  \frac{s_{\mathrm w}^2}
                                         {c_{\mathrm w}^2-s_{\mathrm w}^2} 
\De r^{\rm SUSY} .
\label{deltaMW}
\end{equation}
Here $\De r^{\rm SUSY}$ represents the one-loop contribution from the
supersymmetric particles of the considered sector of the MSSM and
$\MW^{\rm ref} = 80.425 \gev$, see \citere{MWweber} for more details.

\begin{figure}[htb!]
\begin{center}
\includegraphics[width=12cm,height=6cm]{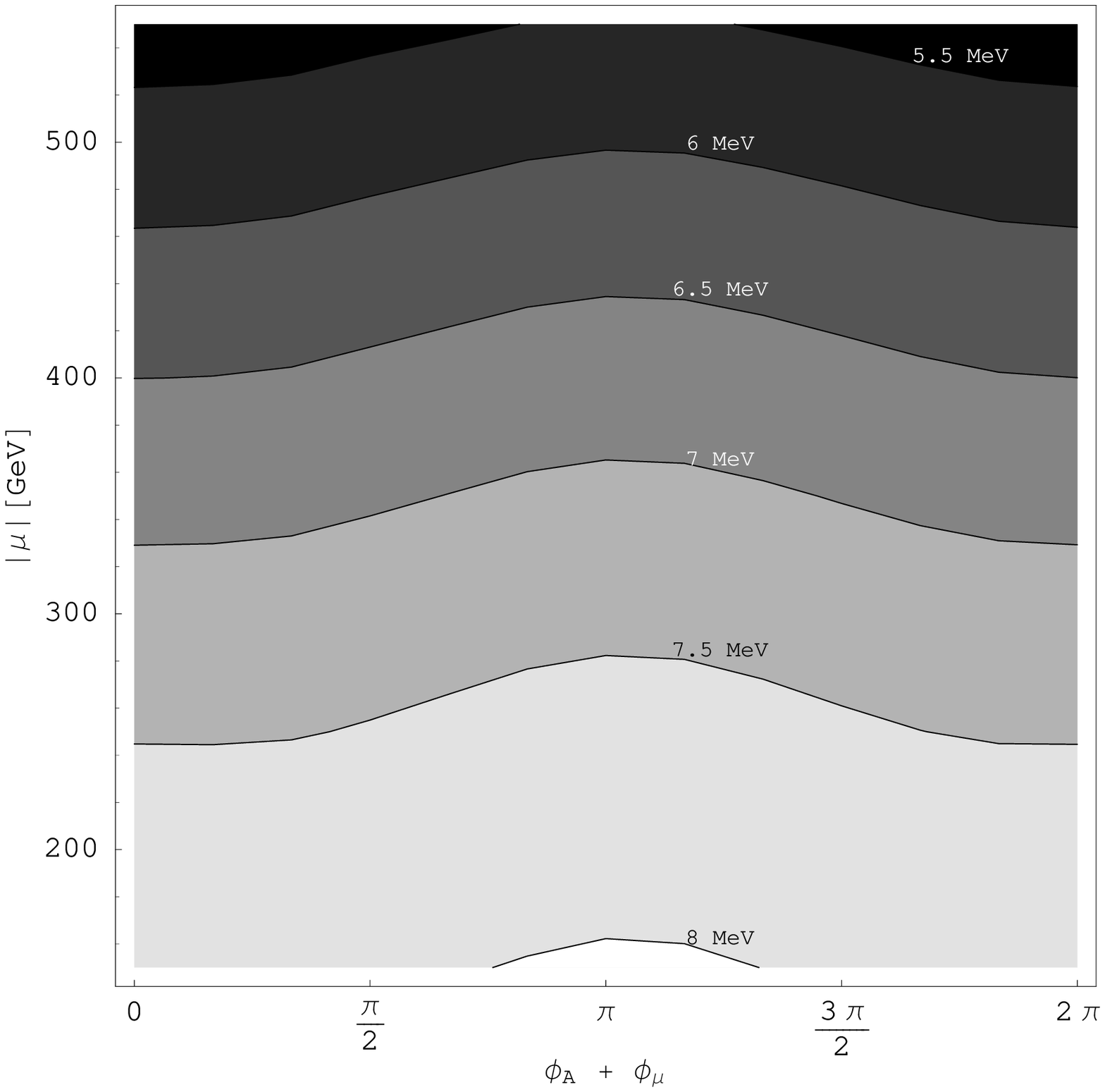}
\caption{Squark  contributions to $\de \MW$  as function of the phase
 $(\phi_A + \phi_\mu)$, where $\phi_A \equiv \phi_{\At} = \phi_{\Ab}$,
for different values of the common sfermion mass $\msusy=500$, $600$,
 $1000$ GeV.
The other relevant SUSY parameters are set to  $\tb = 5$, 
$|A_{t,b}|=2\msusy$, $|\mu|=900$ GeV.} 
\label{deltaMWSfermMUEPhase}
\end{center}
\end{figure}

\begin{figure}[htb!]
\begin{center}
\includegraphics[width=5.8cm,height=5cm]{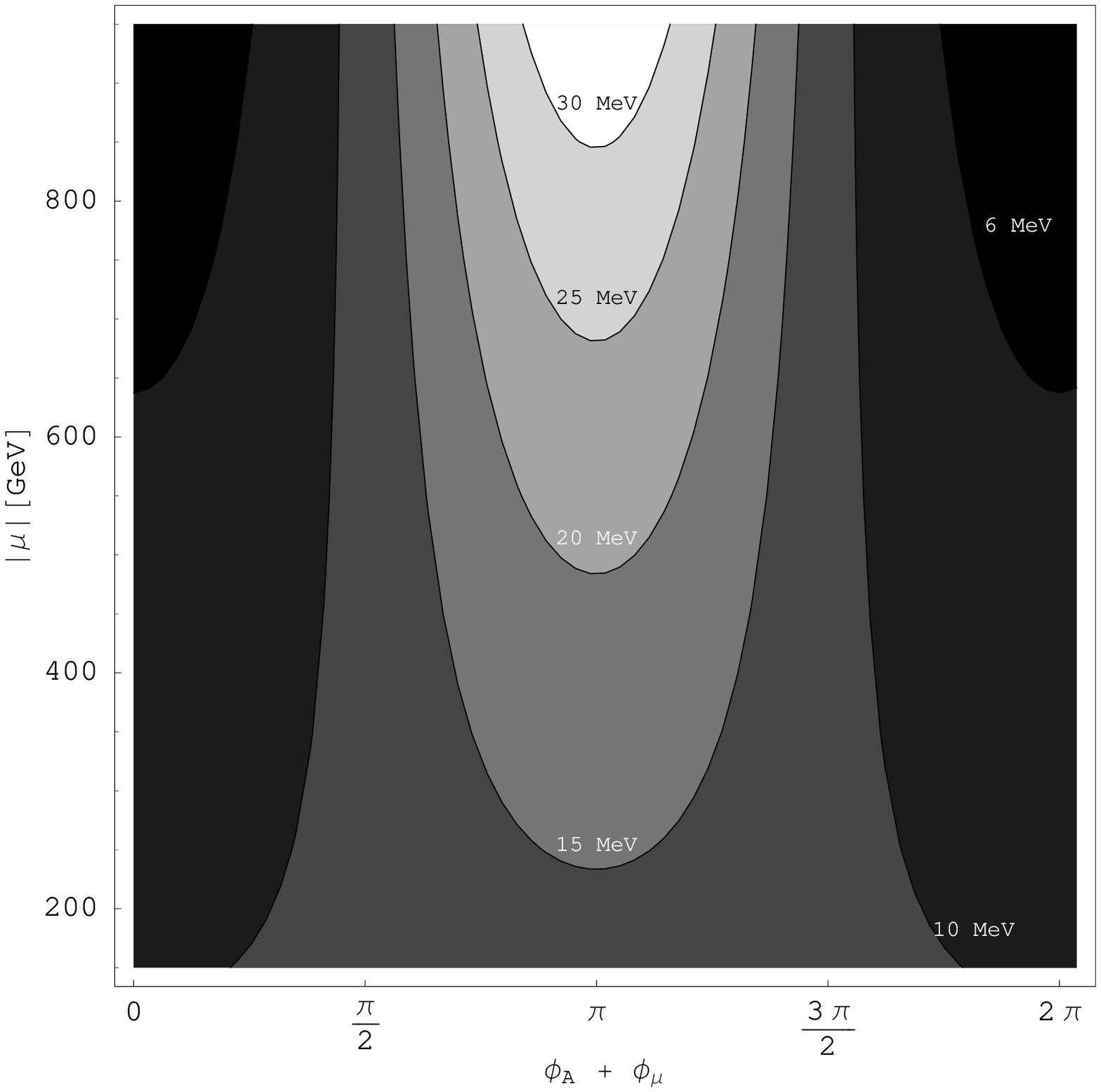}
\hspace{.2cm}
\includegraphics[width=5.8cm,height=5cm]{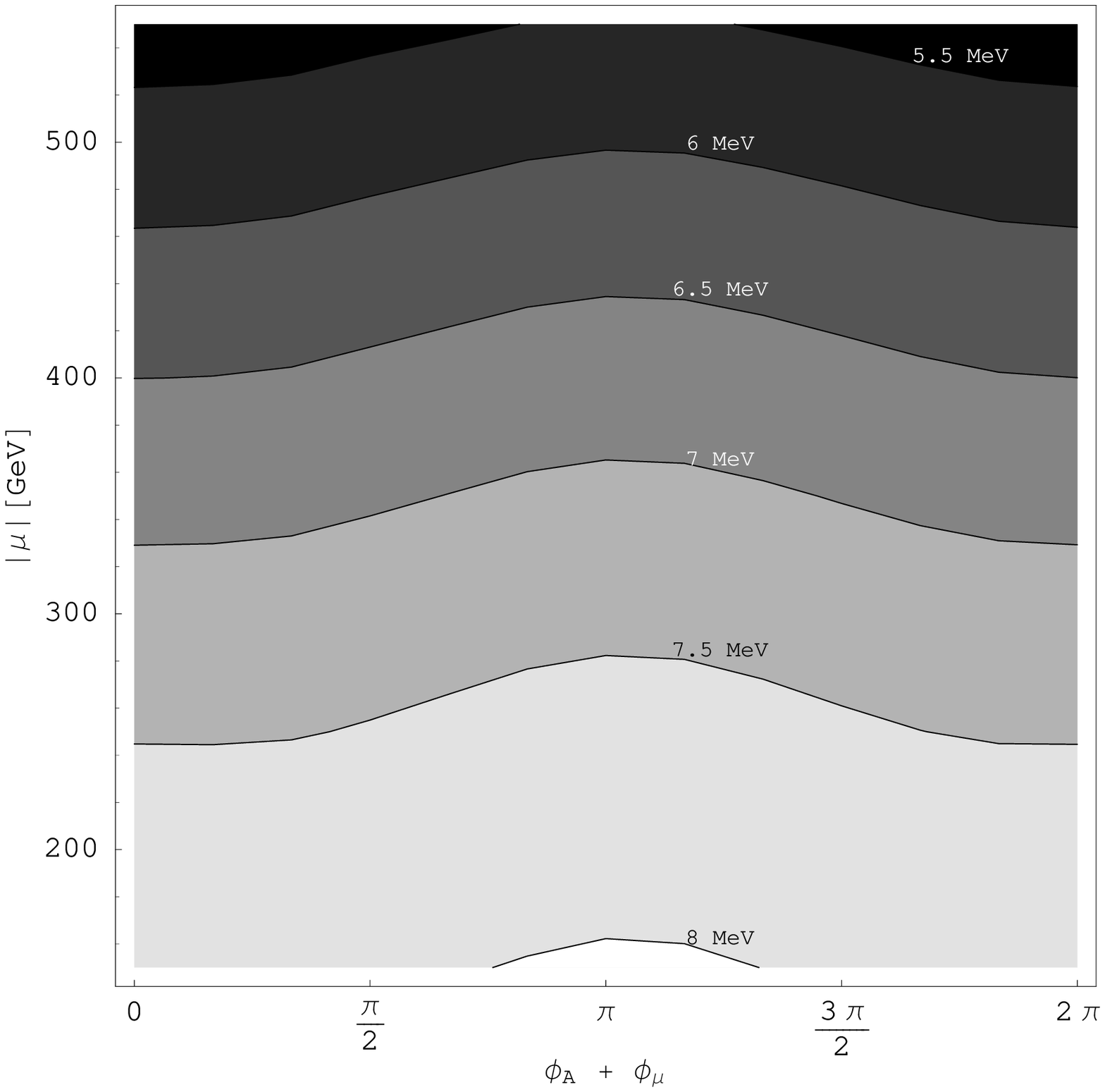}
\caption{Contour lines of the squark contributions to 
$\de\MW$ in the plane of $(\phi_A + \phi_\mu)$ 
and $|\mu|$, where $\phi_A \equiv \phi_{\At} = \phi_{\Ab}$.
The left plot shows a scenario with $\tb = 5$, $\msusy = 500 \gev$,
$|A_{t,b}|=1000 \gev$, while in the right plot $\tb = 30$, 
$\msusy = 600 \gev$, $|A_{t,b}|=1200 \gev$.}
\label{deltaMWMMUEphase}
\end{center}
\end{figure}

The leading one-loop SUSY contributions to $\De r$ arise
from the $\Stop/\Sbot$ doublet. 
The complex parameters in the $\Stop/\Sbot$ sector are
$\mu$, $\At$ and $\Ab$, entering via the off-diagonal entries of the
$\Stop$~and $\Sbot$~mass matrices, $X_{t,b}$. 
In \citere{MWweber} it has been shown at the analytical level that the phases
$\phi_{X_{t,b}}$ drop out entirely in the full one-loop calculation of
$\De r$ and have no influence on $\MW$. Hence, the phases and absolute
values of $\mu$, $\At$ and $\Ab$ enter the sfermion-loop
contributions (at one-loop order) only via a shift in the $\Stop$~and
$\Sbot$ masses and mixing angles.

The phase dependence is illustrated in 
\reffis{deltaMWSfermMUEPhase} and \ref{deltaMWMMUEphase}.
\reffi{deltaMWSfermMUEPhase} shows the 
effect on $\de\MW$ from varying the phase $(\phi_A+\phi_\mu)$ for a
fixed value of $|\mu|=900 \gev$ and $\msusy=500$, $600$, $1000 \gev$,
while in \reffi{deltaMWMMUEphase} the squark sector contributions to
$\de\MW$ are shown as contour lines in the plane of $(\phi_A+\phi_\mu)$
and $|\mu|$. In the scenario with $\tb=5$ (\reffi{deltaMWSfermMUEPhase}
and left panel of \reffi{deltaMWMMUEphase}) the variation of 
$(\phi_A+\phi_\mu)$ can amount to a shift in $\MW$ of
more than $20 \mev$. The most pronounced phase dependence is obtained
for the largest sfermion mixing, i.e.\ the smallest value
of $\msusy$ and the largest value of $|\mu|$.
The right panel of \reffi{deltaMWMMUEphase} shows a scenario with
$\tb=30$. The plot clearly displays the resulting
much weaker phase dependence compared to the scenario in the left panel of
\reffi{deltaMWMMUEphase}.
The variation of the complex phase gives rise
only to shifts in $\MW$ of less than $0.5$~MeV, while changing $|\mu|$
between $100$ and $500$~GeV leads to a shift in $\MW$ of about 2~MeV.


\section{The MSSM parameter scan}
\label{sec:scan}

Here we show the overall behaviour of $\MW$ in the MSSM by
scanning over a broad range of the SUSY parameter space. All relevant
SUSY parameters are varied independently of each other, see
\citere{MWweber} for details.
We have taken into account the constraints on the MSSM parameter space
from the LEP Higgs searches~\cite{LEPHiggsMSSM,LEPHiggsSM} and the lower
bounds on the SUSY particle masses from \citere{pdg}.

\begin{figure}[htb!]
\begin{center}
\includegraphics[width=12cm,height=8.5cm]{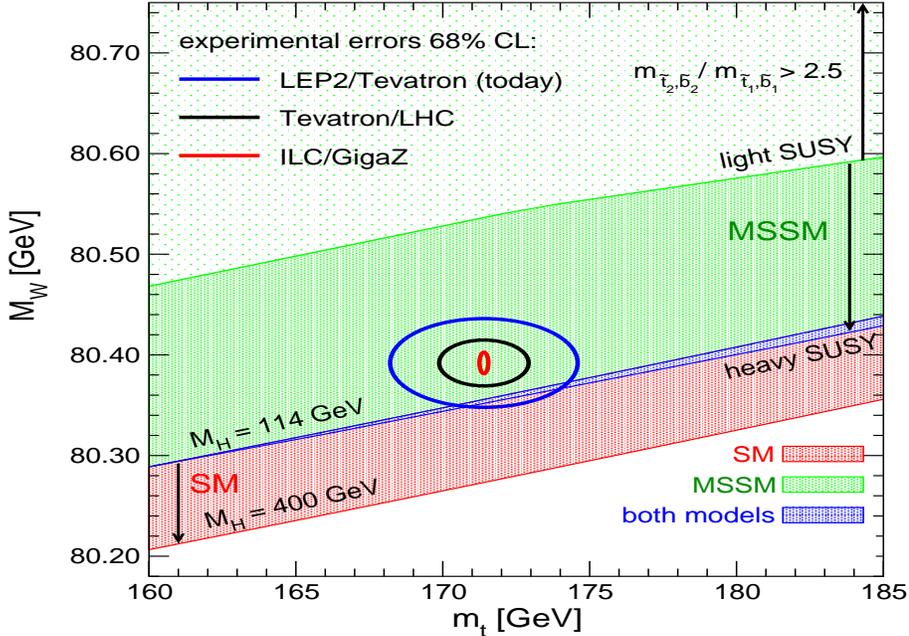}
\begin{picture}(0,0)
\CBox(-110,30)(-15,43){White}{White}
\end{picture}
\caption{Prediction for $\MW$ in the MSSM and the SM as a function of
$\mt$ in comparison with the present experimental results for $\MW$ and
$\mt$ and the prospective accuracies (using the current central values)
at the Tevatron / LHC and at the ILC. 
Values in the very light shaded region can only be
obtained in the MSSM
if at least one of the ratios $\mstz/\mste$ or $\msbz/\msbe$
exceeds~2.5.}
\label{fig:MWMT}
\end{center}
\end{figure}

In \reffi{fig:MWMT} we 
compare the SM and the MSSM predictions for $\MW$
as a function of $\mt$ as obtained from the scatter data (the plot
shown here is an update of \citeres{PomssmRep,MWweber,MWMSSM1LA}).
The predictions within the two models 
give rise to two bands in the $\mt$--$\MW$ plane with only a relatively small
overlap region (indicated by a dark-shaded (blue) area in \reffi{fig:MWMT}). 
The very light-shaded (light green), the light shaded (green) and the
dark-shaded (blue) areas indicate 
allowed regions for the unconstrained MSSM. 
In the very light-shaded region 
at least one of the ratios $\mstz/\mste$ or $\msbz/\msbe$ exceeds~2.5.
The current 68\%~C.L.\ experimental results
$\mt^{\rm exp} = 171.4 \pm 2.1 \gev$ and 
$\MW^{\rm exp} = 80.392 \pm 0.029 \gev$ are indicated in the plot. 
As can be seen from 
\reffi{fig:MWMT}, the current experimental 68\%~C.L.\ region for 
$\mt$ and $\MW$ exhibit a slight preference of the MSSM over the SM.




\end{document}